# Initiation and stagnation of room temperature grain coarsening in cyclically strained gold films.

**Oleksandr Glushko*[1], Gerhard Dehm[2]**

[1]*Erich Schmid Institute of Materials Science, Austrian Academy of Sciences, Jahnstraße 12, 8700 Leoben, Austria*
[2] *Max-Planck Institut für Eisenforschung GmbH, Max-Planck-Straße 1, 40237 Düsseldorf, Germany*

*corresponding author: oleksandr.glushko@oeaw.ac.at

**Abstract.** Despite the large number of experiments demonstrating that grains in a metallic material can grow at room temperature due to applied mechanical load, the mechanisms and the driving forces responsible for mechanically induced grain coarsening are still not understood. Here we present a systematic study of room temperature grain coarsening induced by cyclic strain in thin polymer-supported gold films. By means of detailed electron backscatter diffraction analysis we were able to capture both the growth of individual grains and the evolution of the whole microstructure on the basis of statistical data over thousands of grains. The experimental data are reported for three film thicknesses with slightly different microstructures and three different amplitudes of cyclic mechanical loading. Although different kinds of grain size evolution with increasing cycle number are observed depending on film thickness and strain amplitude, a single model based on a thermodynamic driving force is shown to be capable to explain initiation and stagnation of grain coarsening in all cases. The main implication of the model is that the grains having lower individual yield stress are coarsening preferentially. Besides, it is demonstrated that the existence of local shear stresses imposed on a grain boundary is not a necessary requirement for room-temperature grain coarsening.

## 1. Introduction.

It has been well established that mechanical stress can assist or even induce migration of grain boundaries (GBs) in metallic materials. A clear evidence of mechanically assisted GB migration was provided in bicrystal experiments where macroscopic displacement of a flat GB due to applied shear stress was continuously observed using optical microscopy [1–3]. It was demonstrated that not only low-angle GBs, which can be represented as aligned rows of dislocations, but also high-angle GBs can migrate under applied mechanical stress. With the help of molecular dynamics simulations it was shown that migration of different GBs can occur if normal GB motion is accompanied by shear displacement of the grain behind the migrating boundary [4–6]. This phenomenon is generally called shear-coupled GB migration, although there is a number of different models predicting the relationship between normal

GB translation and shear deformation [5–8]. Shear coupling is described by the coupling factor which is defined as a ratio of shear displacement behind the migrating boundary to the normal displacement of GB position. The atomistic origin of shear-coupled GB migration lies in the existence of GB steps with dislocation core structure, also called disconnections. First experimental observations of GB dislocations were reported in 1980s during in-situ hot stage transmission electron microscopy (TEM) [8–10]. More recent in-depth experimental investigations revealed that movement of disconnections results in normal displacement of the GB. The amount of shear displacement behind a migrating GB depends on the Burgers vectors which can be assigned to the dislocation core of active disconnection [11–13]. Therefore, even the same GB can migrate in different modes resulting in positive, negative or zero coupling factor, depending on the structure of the dislocation core of active disconnections. Although shear coupling during GB migration is well confirmed by a number of theoretical [4–6,14] and experimental studies [7,11–13,15–17], it is still unclear how general this process is. Even in bicrystal experiments, i.e. when shear displacement of the grains is not constrained by the surroundings, it was shown that that GBs can migrate over long distances without shear coupling [18]. In polycrystalline materials shear displacement of the grain behind the migrating boundary must be significantly constrained by the neighboring grains [19].

Mechanically induced grain coarsening has been observed in a large variety of materials at room temperature [20–32]. Although GB migration and grain coarsening are closely related and are often mentioned in the same context, it is important to highlight an important difference between these two processes. Let us consider three grains (G1, G2, G3) separated by two GBs (GB1, GB2) and assume the G3 grows in expense of G2 while the third grain (G1) remains unchanged, as shown in Fig. 1. Migration of GB2 towards GB1 (Fig. 1a) results in collapse of the grain 2. As soon as grain 2 disappears, both GB1 and GB2 disappears as well, and a new grain boundary GB* (Fig. 1b) which did not exist in the initial microstructure is created. Thus, grain coarsening is a two-step process consisting of GB migration and GB elimination and leads not only to an increase of average grain size, but also to permanent elimination of "old" GBs and appearance of new GBs. It is also important to note that the total free energy of the system may remain unchanged if only GB migration occurs while in the case of grain coarsening the total free energy always decreases by the amount of the energy associated with eliminated GBs.

Despite the large number of scientific reports, the mechanisms and driving forces leading to collective mechanically induced grain coarsening remain unknown. The main findings can be summarized by rather qualitative conclusion that the grain coarsening is observed within the areas with high local mechanical stress [22,23,25,27,28,30].

Some attempts to describe the grain coarsening effect in more quantitative way were recently performed by in-situ TEM straining experiments employing crystal orientation mapping [22,23]. In both these works relatively small areas containing only few tens of grains located close to stress concentration points were examined. Due to the missing statistics, the driving forces for grain coarsening remained uncovered and no explanation, why some grains do coarsen and others not, was provided. A different approach involving polymer-supported ultra-fine grained (UFG) films and utilizing cyclic straining in combination with electron backscatter diffraction (EBSD) analysis allowed to monitor the evolution of about ten thousand grains [33] with increasing number of straining cycles. By quasi in-situ EBSD evaluation it was demonstrated that coarsened grains neither share a common orientation with respect to applied strain nor possess similarities in GB properties. However, it was found that initially larger grains have higher chance to coarsen during cyclic loading. To explain this, a simple thermodynamic driving force model was proposed, although without a proof for its generality since only one thin film system (500 nm thick Au film) was analyzed.

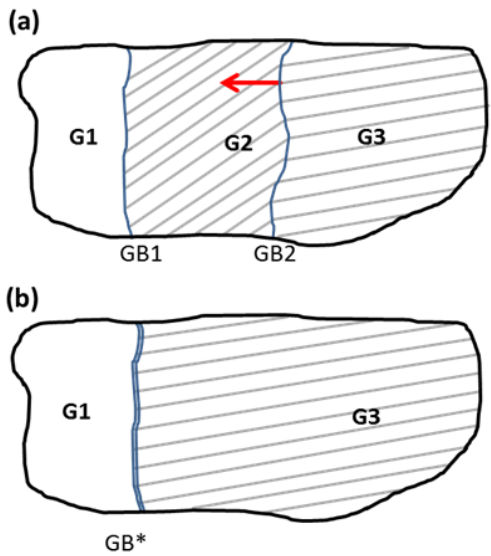

Fig. 1. Schematic representation of grain coarsening mediated by grain boundary migration. Three grains (G1, G2, G3) separated by two GBs (GB1, GB2) are shown in (a). Successive migration of GB2 in the direction of the arrow finally leads to the situation shown in (b) where G2, GB2 and GB1 disappear and a new GB* appears.

In the current work the room-temperature grain coarsening is studied systematically involving three different film thicknesses and three different amplitudes of applied mechanical strain. By combining quasi in-situ and post-mortem EBSD characterization significant statistical data about the evolution of grain sizes with increasing number of straining cycles was collected and evaluated. It will be demonstrated that the thermodynamic driving force model explains well not only the initiation of grain coarsening but also its stagnation.

## 2. Experimental details.

The gold films with the thicknesses of 250 nm, 500 nm, and 1000 nm were deposited on 50 μm polyimide UPILEX S substrates by electron beam evaporation in a Balzers BAK 550 evaporation machine with the vacuum of $2.1 \times 10^{-7}$ mbar and using a deposition rate of 0.3 nm/s. The deposition was performed at room temperature. No annealing or any other post-treatment procedures were applied to the films. The test samples with the width of 4 mm and length of 40 mm were cut using a scalpel out of larger sheets. Mechanical testing was performed on MTS Tytron 250 tensile testing device using a gage length of 20 mm. Cyclic loading was applied in strain control mode by means of a sine strain function oscillating with the frequency of 0.5 Hz between the initial position (zero strain) and a defined value of peak strain. The values of peak strains (0.5%, 1%, 2%) were assigned to the tensile testing device by defining the peak displacements (100 µm, 200 µm, and 400 µm, respectively). The tensile testing device employed internal extensometer to reach the assigned displacements cyclically by applying a strain-time sine function with the frequency of 0.5 Hz. In order to evaluate the effect of machine stiffness, the strain was also measured directly on the sample in a non-contact manner by correlating photo images of the sample surface under different values of nominal strain. It was found that the sample strain is approximately 5% less than the nominal strain calculated from the displacement of the machine arm (i.e. 1.9% instead of 2%). Since this difference does not make any changes to the manuscript conclusions it was decided to refer to the values of nominal strain for clarity. The pre-strain was in the range 0.02%-0.05% corresponding to arm displacements up to 10 µm. After application of a given number of cycles with a given value of peak strain each sample was characterized using electron backscatter diffraction analysis (EBSD). EBSD images were taken using EDAX system with OIM analysis™ in combination with scanning electron microscope (SEM) Zeiss Leo 1525 using 20 kV acceleration voltage, 120 µm aperture, working distances between 9 mm and 11 mm and the step sizes between 40 and 60 nm. Prior to each EBSD characterization the vacuum chamber of the SEM was cleaned for approximately 15 min by plasma cleaner to reduce e-beam induced carbon contamination of the surface. Typical surface areas analyzed by EBSD in each sample were between 300 µm$^2$ for as deposited films (containing 10000-20000 grains) and up to 1500 µm$^2$ for the films after grain coarsening (containing 3000-8000 grains). The only post-processing procedure which was applied to the EBSD data was removing of scan points which have the confidence index less than 0.05. These points appear in black in all EBSD scans shown in this paper. Note that the grains smaller than 30-40 nm cannot be clearly resolved by EBSD technique since this length scale is close to the typical sizes of interaction volume of incident electron with the crystal lattice. A set of samples was used for quasi-in-situ EBSD characterization in order to reveal grain growth evolution. Focused ion beam (FIB) milled markers were introduced in several areas of those samples and EBSD analysis was performed prior to, at intermediate straining cycles, and after the mechanical test.

## 3. Results.

### 3.1. Initial microstructure.

The initial grain size distributions of the three types of gold films are given in Fig. 2. The 250 nm thick films have log-normal grain size distribution with average values of 145±40 nm. The grain size distribution in 500 nm thick films is also close to log-normal but with a "tale" indicating a small fraction of larger grains. The average grain size of 500 nm thick films is 210±60 nm. These two film types exhibit very strong fiber texture with about 90% of grains having (111) orientation in surface normal direction within maximum tolerance misorientation of 10°. As follows from Fig. 2, the 1 µm thick films have a more pronounced tale in the grain size distribution which can be also considered as a weak bi-modal distribution. Although most of the grains have sizes around 200 nm, there are some larger grains with the sizes up to 1 µm. The initial average grain size for 1 µm thick gold films is calculated to be 320±80 nm. The texture of 1 µm thick films is less pronounced with only about 50% grains with (111) orientation within a maximum tolerance angle of 10°. The description of the initial microstructure of the three sample types is given in much more details in the Supplementary Fig. S1. The initial microstructure of the films was proven to be stable by comparing the analysis performed a few weeks after the deposition and one year after the deposition.

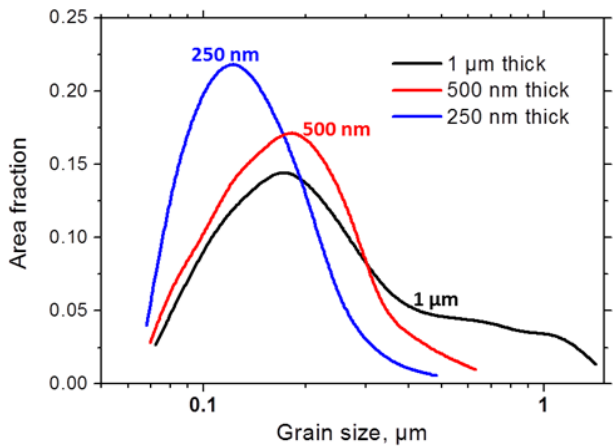

Fig. 2. Area-weighted grain size distributions of the initial microstructure of three gold films under investigation.

## 3.2. Quasi in-situ characterization of the grain size evolution in 1 µm thick gold films.

In order to capture the evolution of the grain size with the cycle number a set of tests with quasi-in-situ EBSD characterization was performed as described in the experimental section. In Figs. 3(a-d) the grain orientation maps in TD direction, i.e. in the direction parallel to the applied strain, of the same area of 1 µm thick Au film after 0, 500, 1000, and 5000 cycles with 1% cyclic strain are shown. The corresponding SEM images are given in Figs. 3(e-h) and the grain size distributions taken from the same area are shown in Figs. 3(i-l). The dashed circle in Figs. 3a and 3e depict a FIB marker which was used to align the sample for imaging.

By comparing the microstructure shown in Fig. 3a and Fig. 3b one can clearly see that significant grain coarsening occurs already during the first 500 cycles. It is important to notice that the vast majority of the coarsened grains were initially larger, i.e. they belong to the "tale" of the initial grain size distribution (marked in Fig. 3i). After 5000 cycles the average grain size calculated from the same area reaches 1860±420 nm. The black spots which appear in the grain orientation map after 5000 cycles are associated with local surface roughening which leads to insufficient quality of the electron diffraction pattern. The SEM images of the surface demonstrate that after 500 and 1000 cycles there are only few visible slip steps and slip bands predominantly within the larger grains. After 5000 cycles (Fig. 3f) pronounced slip bands and extrusions are clearly visible. As follows from the grain size distribution plots (Figs. 3i-3l) the initial peak at about 200 nm gradually decreases with the cycle number and totally disappears after 5000 cycles. Application of further straining cycles did not lead to further grain coarsening but only to further development of fatigue damage in the form of more extensive slip bands, growing extrusions, and cracks, as shown below in Fig. 5.

The following remarks can be made on the basis of the quasi-in-situ EBSD study. First of all, by comparing Figs. 3a and 3d it is clear that there are virtually no areas which maintained the original microstructure. This means that all grain boundaries independently from their misorientation angle, type or symmetry have migrated during cyclic loading. Second, there are no preferable crystallographic orientations among the coarsened grains. Third, there is no preferable direction of the grain extension although the applied strain was in horizontal direction. As one can see the grains grow in all directions, i.e. isotropically. Forth, although most of the coarsened grains were initially larger than the average, one can find several exceptions out of this trend. Four examples of grains which were too small to be clearly detectable in the initial microstructure but grew substantially during 5000 cycles are marked by white arrows in Fig. 3d. It is shown in the Supplementary figure Fig. S2 that most of these grains exhibit a low Taylor factor.

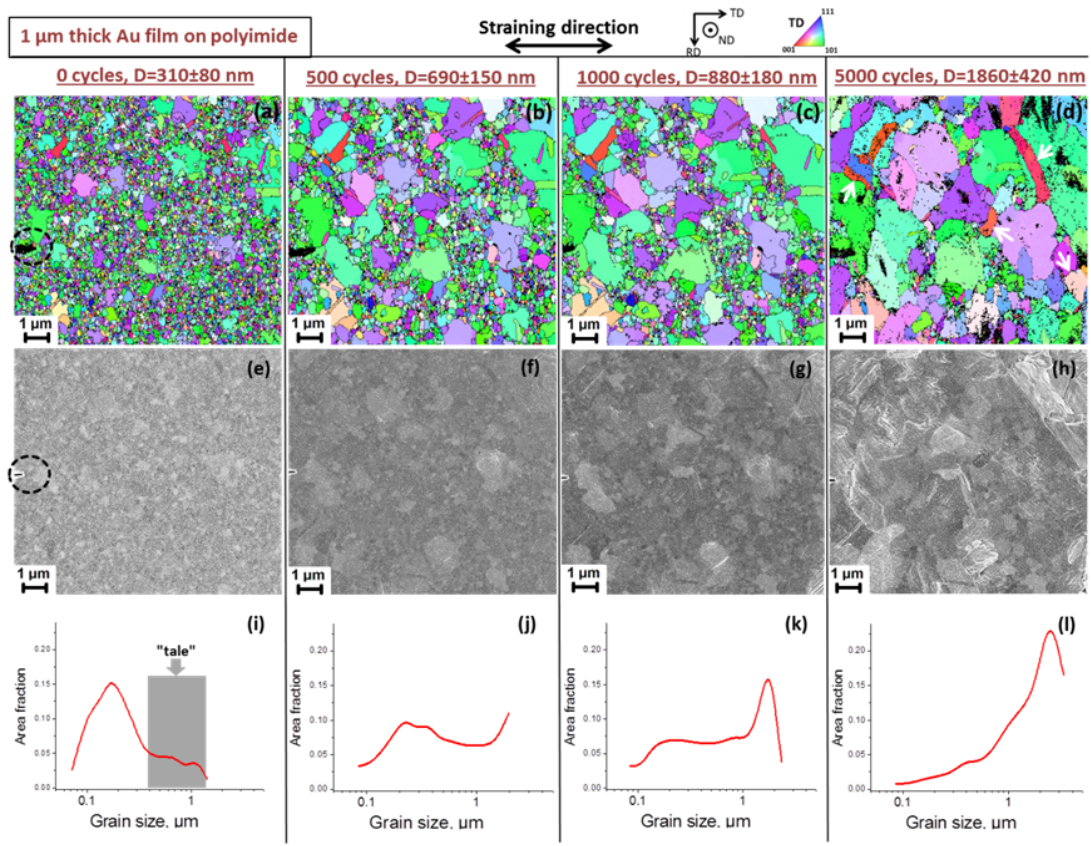

Fig. 3. Grain orientation maps in TD direction (parallel to the applied strain) for the same area of 1 μm thick Au film in as-deposited state (a), after 500 cycles (b), after 1000 cycles (c), and after 5000 cycles (d) with 1% peak strain. The corresponding SEM images are shown in (e)-(h). The corresponding grain size distributions are shown in (i)-(l). The dashed circles in (a) and (e) depict the tip of a FIB-milled marker used to find the same area after each portion of straining cycles. Four white arrows in (d) point at the grains which were not clearly detectable in the initial microstructure but became relatively large during cyclic straining. The straining direction is horizontal.

### 3.3. Collective grain coarsening effect.

To investigate the collective effect of grain coarsening the three gold film types were subjected to cyclic strain of three different amplitudes, as described in the experimental section. The evolution of the average grain size with the cycle number is shown in Fig. 4. Each point on the plots corresponds to a separate sample loaded to the corresponding cycle number without interruption. The error bars show the standard deviations calculated using EBSD data from surface areas in the range between 300 µm$^2$ and 1200 µm$^2$. It should be kept in mind that characterization of grain sizes by means of average value and standard deviation may be not totally correct if the grain size distribution is non-Gaussian, but for instance bi-modal. Nevertheless it was used in order to provide a consistent description of the grain size evolution.

The 1 µm thick films loaded to cyclic strain of 2% demonstrate very fast growth of the grain size with the cycle number with subsequent saturation after 2000 cycles (Fig. 4a). The grain size at saturation is about 2 µm. For 1% cyclic strain the grain coarsening with the cycle number occurs somewhat slower and the saturation is achieved after 5000 cycles. For the cyclic strain of 0.5% gradual increase of the grain size with the cycle number is observed without saturation up to 30000 cycles.

In the case of 500 nm thick films the grain evolution is similar for the cyclic strains of 1% and 2% (Fig. 4b). The grain size at saturation is between 1.5 µm and 2 µm. However, for the cyclic strain amplitude of 0.5% only single, sparsely distributed grains have coarsened up to the sizes of few micrometers while the rest of the grains remained unaffected. The bi-modality of the grain size distribution after 30000 cycles with 0.5% strain is depicted by the letters "bm" in Fig. 4b.

The 250 nm thick films demonstrate slower grain coarsening with the cycle number. The cyclic strain of 0.5% did not induce any detectable grain coarsening in the film. Clear saturation of grain sizes is observed for the strain amplitude of 2% while in the case of 1% cyclic strain the saturation, most probably, is not reached after 30000 cycles. Interestingly, bi-modal grain size distributions are observed in all 250 nm thick films where the grain coarsening was initiated.

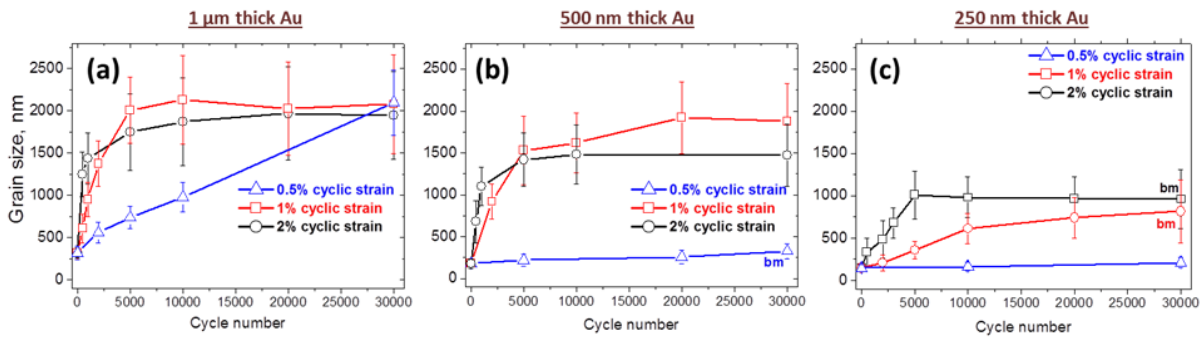

Fig. 4. Evolution of the grain size with the cycle number for (a) 1 µm thick films, (b) 500 nm thick films, and (c) 250 nm thick films. The characters "bm" in (b) and (c) point that the corresponding grain size distributions are strongly bi-modal after 30000 cycles.

As follows from Fig. 4, there are three types of the observed grain size evolution observed up to 30000 cycles: (i) growth and saturation, (ii) continuous growth, and (iii) minor or negligible growth. Such, rather complex, behavior leads to a natural question of whether it is possible to explain it in the framework of a single model?

### 3.4. Microstructure and surface damage.

Apart from the grain coarsening, cyclic mechanical loading induces fatigue damage which can appear in the form of slip steps, extrusions and cracks. The SEM images of three film thicknesses exposed to 30000 cycles with 1% strain are shown in Fig. 5. These images were taken using the in-lens detector as it provides better image quality at high magnifications. The brighter areas of the images correspond to the fatigue-induced surface features like slip bands and extrusions. The parts of the surface which were not affected by the cyclic loading appear in dark gray. As one can see in Fig. 5a-5d both the 500 nm thick film and 1 µm thick film exhibit heavily damaged surface with numerous slip bands, massive extrusion and cracks. Only small fractions of the surface area remained undamaged.

A tremendously different picture is observed for the 250 nm thick films (Fig. 5e-5f). Here one can see single slip steps and extrusions formed within individual, heavily coarsened grains. The rest of the surface exhibits no visible fatigue damage. One can also see a small crack initiated inside a grain in Fig. 5f. It will be shown in the section 4.2 that application of further mechanical loading leads to propagation of the cracks initiated from such individual coarsened grains.

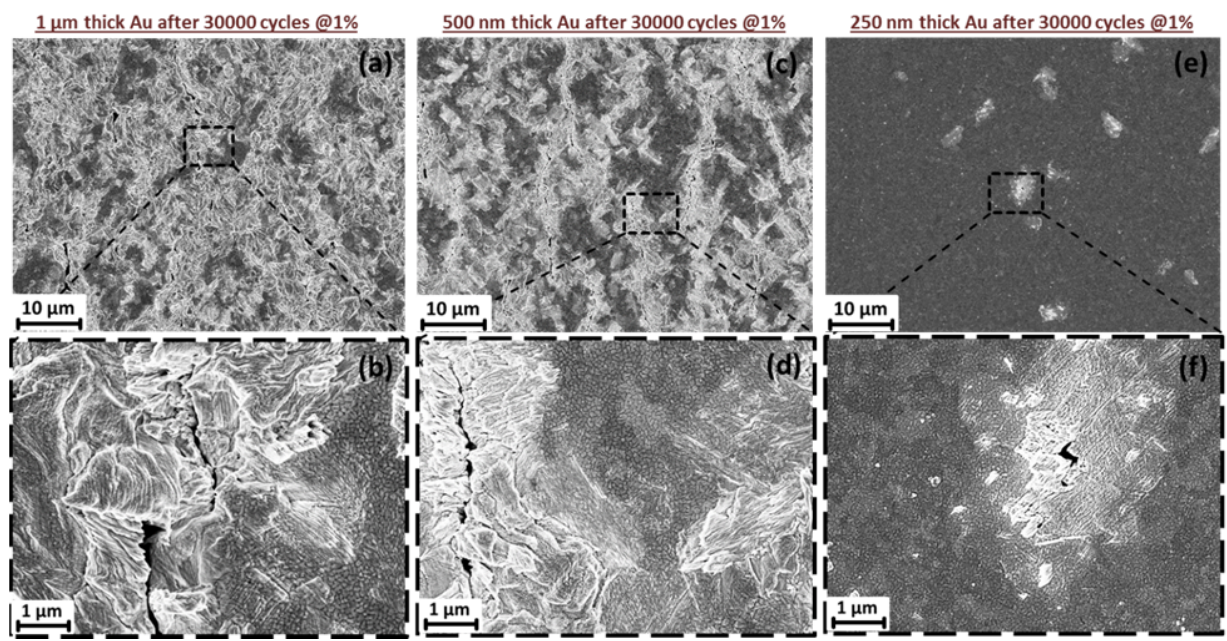

Fig. 5. Scanning electron microscopy images of the surface of 1 µm (a,b), 500 nm (c,d), and 250 nm (e,f) thick gold films after 30000 cycles with 1% peak strain. Images (b,d,f) show magnified areas depicted by dashed boxes in (a,c,e), correspondingly. The straining direction is horizontal.

# 4. Discussion.

## 4.1. The driving force for the grain coarsening.

The observed grain coarsening can be explained well by the recently proposed thermodynamic driving force model [33]. This model suggests that some grains which have lower individual yield stress start to deform plastically with the increasing cycle number earlier than the rest. Local plastic deformation results in local stress relaxation which, in turn, leads to the lower elastic strain energy density within these individual grains. The difference in apparent elastic strain energy density between the individual grains provides a thermodynamic driving force which can induce the migration of GBs and grain coarsening. This model can be visualized by a simple schematic shown in Fig. 6. Before mechanical strain is applied the free energy density can be assumed to be virtually equal in neighboring grains (Fig. 6a). This assumption is justified by the fact that no time-dependent change in the microstructure of the films was observed. In mechanically loaded state the free energy density in each grain is increased by the amount of elastic strain energy density, $\Delta E$. If the grain 3 (G3) has yielded and the grain 2 (G2) deformed purely elastically, then a thermodynamic driving force $\Delta P = \Delta E_{G2} - \Delta E_{G3}$ appears and the grain G3 is driven to grow in the expense of grain G2. In the sketch shown in Fig. 6 it is assumed that the GB between grain 1 and grain 2 does not migrate due to the insufficient driving force. To summarize, this model predicts that grains which have lower strain energy density in the strained state should coarsen preferably under cyclic mechanical loading. The results of quasi-in-situ EBSD analysis shown in Fig. 3 correspond very well to the predictions of the thermodynamic driving force model. The vast majority of the coarsened grains were already initially larger than the surrounding grains which means that they have lower yield stress according to the Hall-Petch effect. Some of the coarsened grains in Fig. 3d were not initially larger but have very low Taylor factor (supplementary Fig. S2) which also means that plastic slip occurs in these grains more readily than within the surrounding grains.

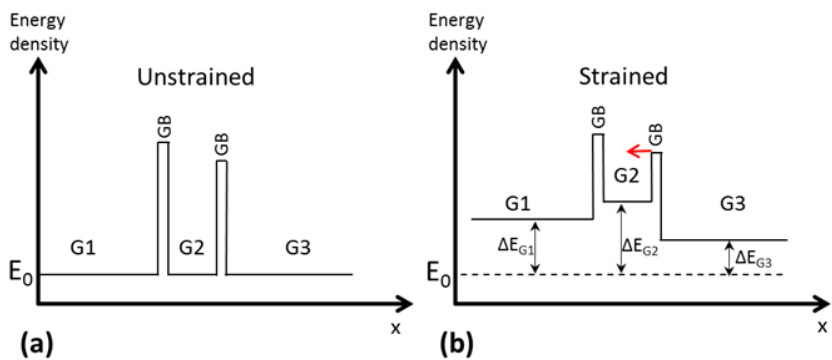

Fig. 6. Appearance of the thermodynamic driving force under applied mechanical strain. In unstrained state (a) different grains (G1, G2, G3) have virtually the same free energy density ($E_0$) and no grain boundary migration is observed. Mechanical strain (b) increases the free energy density of individual grains by the amount of elastic strain energy ($\Delta E_{G1}$, $\Delta E_{G2}$, $\Delta E_{G3}$). Now a thermodynamic driving force exists for the growth of the grain with lowest elastic strain energy density (G3) in expense of the grain with higher elastic strain energy density (G2).

The thermodynamic driving force model can also explain the difference in collective grain coarsening at the cyclic strain of 0.5% shown in Fig. 4. The 250 nm thick films show no grain coarsening at 0.5% cyclic strain which suggests that all, even the largest, grains stay within the purely elastic regime. In the case of 1 µm thick films the grain coarsening occurs but the average grain size grows significantly slower with the cycle number in comparison to higher cyclic strains. One can suppose that coarsening was activated only within the large grains from the tale of the grain size distribution (Fig. 2) which had high enough difference in yield strength in comparison to neighboring grains. The 500 nm thick films loaded at 0.5% cyclic strain stay mostly elastic with the exception of the largest grains which coarsened during 30000 cycles leading to strongly bi-modal grain size distribution.

Although the activation of the grain coarsening can be explained by the thermodynamic driving force, there are two open questions which need to be discussed. Firstly, why does the grain coarsening stagnate? A clear saturation of grain sizes is observed in 1 µm and 500 nm thick films for 1% and 2% strain amplitude and in 250 nm thick films for 2% strain amplitude (Fig. 4). Secondly what is the atomistic mechanism of the observed grain coarsening? Although it is not possible to observe the atomistic structure of a grain boundary using EBSD technique, one can discuss how the existing models of GB migration mechanisms could be employed to understand the observed grain coarsening.

### 4.2. The mechanisms causing stagnation of grain coarsening.

The phenomenon of grain size stagnation is still not completely understood even in the case of thermally induced grain coarsening. In a systematic and in-depth study [34] it was concluded that the final grain size distribution obtained for a given material and given annealing temperature cannot be explained by a single driving force. Rather a combination of different driving forces and additional factors, such as grain boundary grooving or impurity drag, must be responsible for stagnation of grain coarsening. It is important to note that during annealing at constant temperature the system can be expected to reach a metastable thermodynamic equilibrium state after a given annealing time. In the case of room-temperature strain-induced grain coarsening considered in the present study the situation is apparently more complex since the grain size evolution is recorded depending on the number of straining cycles. Cyclic mechanical load leads to permanent generation and movement of dislocations which results in formation of slip steps and more developed surface damage in the form of extrusions and cracks with increasing cycle number. Thus, for the grain coarsening induced by cyclic strain no thermodynamic equilibrium is achieved even when the saturation of grain coarsening is observed.

The thermodynamic driving force model predicts that if two grains have similar elastic strain energy density then there is no driving force acting on a GB between them. To understand the *global* stagnation of grain coarsening let us consider the stagnated microstructure after 5000 cycles for 1 µm thick films shown in Fig. 3. It is clearly seen that the initial peak at about 200 nm in the grain size distribution (Fig. 3i) disappears completely after 5000 cycles (Fig. 3l). This can be considered as a topological change of the microstructure. Initially, large "soft" grains are distributed in an UFG "strong" matrix, but at stagnation, UFG "strong" islands are distributed in a large-grained "soft" matrix. We believe that grain coarsening stagnation is caused by this topological inversion and can be explained in the framework of the thermodynamic driving force model with the help of the sketches shown in Fig. 7. In Fig. 7a the initial situation is demonstrated: "soft" grains in a "strong" matrix. In this configuration the matrix reaches higher stress state for a given strain and thus the driving force for grain growth exists (Fig. 7b). Assuming that only the initially larger grains coarsen during the application of mechanical strain, at some point the growing grains will interconnect with each other as shown in Fig. 7c. Now for the same applied strain the maximum stress in the film is limited by the yielding of the "soft" matrix which relaxes through plastic slip and thus the rest of "strong" UFG areas will not reach higher stress state. This means that there will be no significant difference in the strain energy density between the "soft" matrix and "strong" islands and thus the thermodynamic driving force will disappear (Fig. 7d). As one can see, the sketch shown in Fig. 7c corresponds very well to the grain orientation map of stagnated microstructure in Fig. 3d where some UFG islands distributed in heavily coarsened microstructure are still present. It is important to note, that the described stagnation mechanism is somewhat simplified and it does not exclude the existence of local GB migration and grain coarsening even when global saturation is reached. Moreover, in Fig. 7d the effect of the substrate is not considered, thus the distributions of local stresses and strains during loading and unloading portions of each cycle are more complex. Nevertheless, our observations of microstructural evolution always revealed a strong correlation between the grain coarsening stagnation and global interconnection of the coarsened grains.

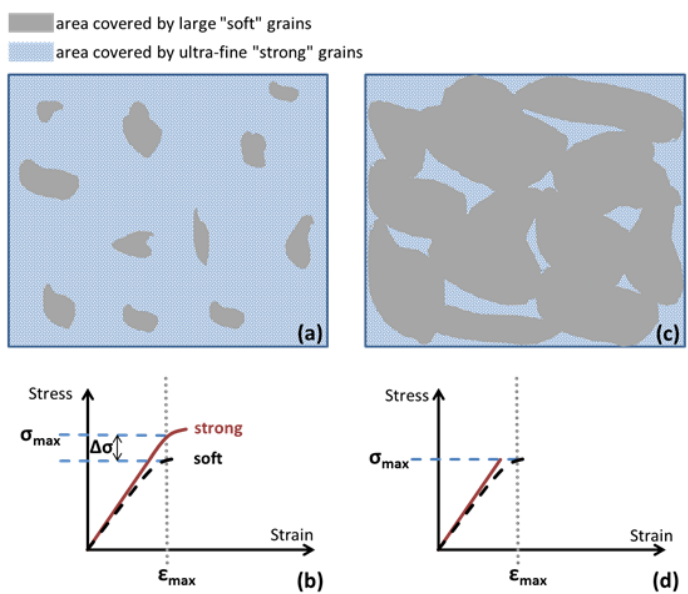

Fig. 7. The mechanism of grain coarsening stagnation due to topological inversion. (a): The initial microstructure can be considered as an UFG matrix with high yield stress and inclusions of larger grains with low yield stress. (b): When exposed to the applied global strain ($\varepsilon_{max}$), the stress in the matrix, which stays in the elastic regime, will be higher than the local stress within the larger grains, which yield. (c): At saturation, the microstructure can be considered as UFG islands distributed in soft matrix consisting of coarsened grains. (d): In such a configuration under the same applied strain ($\varepsilon_{max}$) the maximal stress is restricted by the yield stress of the soft matrix; thus the UFG islands will not reach a higher stress state and the driving force for grain coarsening disappears.

To further confirm that the described topological inversion correlates with the grain coarsening stagnation, additional tests were performed on 1 μm thick films at 0.5% strain since the grain coarsening stagnation is evidently not reached after 30000 cycles (Fig. 4a). The grain orientation maps and the corresponding grain size distributions, provided in the supplementary Fig. S3, confirm that the topological interconnection of the coarsened grains is completed first after 70000 cycles when the saturation of grain coarsening is reached. Interestingly, the saturation grain size of 1 μm thick Au films for 0.5% strain is higher than for 1% and 2% being about 2910±450 nm.

This topological explanation of the grain coarsening stagnation seems not to be applicable for the 250 nm thick films since the microstructure remains strongly bi-modal after 30000 cycles. Further experiments up to 100000 cycles with 1 % strain showed that the average grain size stagnates after 70000 cycles but a strongly bi-modal grain size distribution remains. The SEM image, grain orientation map and grain size distribution of a 250 nm thick gold film after 70000 with 1% cyclic strain are shown in Fig. 8. As follows from Fig. 8a the film exhibits numerous localized crack/extrusion couples running roughly perpendicularly to the applied strain and separated by virtually undamaged areas. It is also clearly visible from the grain orientation map (Fig. 8b) that the coarsened grains are located exclusively along the propagating crack/extrusion couples while the areas between the cracks retain the UFG microstructure. Thus, the first peak in the grain size distribution (Fig. 8c) can be clearly attributed to the areas between the cracks and the second peak corresponds to the coarsened grains along the propagating cracks. It is well-known that formation of a crack in a polymer-supported film results in stress relaxation on the crack edges behind the crack tip as described by the shear lag model [35–37]. Although the shear lag model was developed for brittle coatings under increasing monotonic strain it can be applied qualitatively also here: with increasing mechanical load (in our case – cycle number) a crack propagates until its stress relaxation zone overlaps with the relaxation zone of a neighboring crack. At some point saturation should be reached, i.e. the cracks do not propagate even if further straining cycles are applied. The shortest crack spacing at saturation should be close to the size of the stress relaxation zone while the longest crack spacing can be roughly estimated to be equal to two relaxation zones [37]. The size of this relaxation zone depends on the film and substrate materials, film thickness and strain amplitude [35]. We believe that due to this local stress relaxation the grains between the cracks are preserved from coarsening since they cannot reach high enough stress state and, consequently, high enough driving force does not appear. It is thus to expect that with increasing cycle number only local grain coarsening restricted to the tips of propagating cracks will occur in 250 nm thick films until the crack density reaches the saturation and the stagnated grain structure will remain bi-modal. Thus, the mechanism of grain coarsening stagnation in this case can be described as local stress relaxation due to crack propagation. The very early stage of crack formation was also captured in Fig. 5f were a small crack is visible inside a heavily coarsened grain. It is necessary to mention that localized grain coarsening within the wake of a propagating crack was observed in free-standing films [25,31].

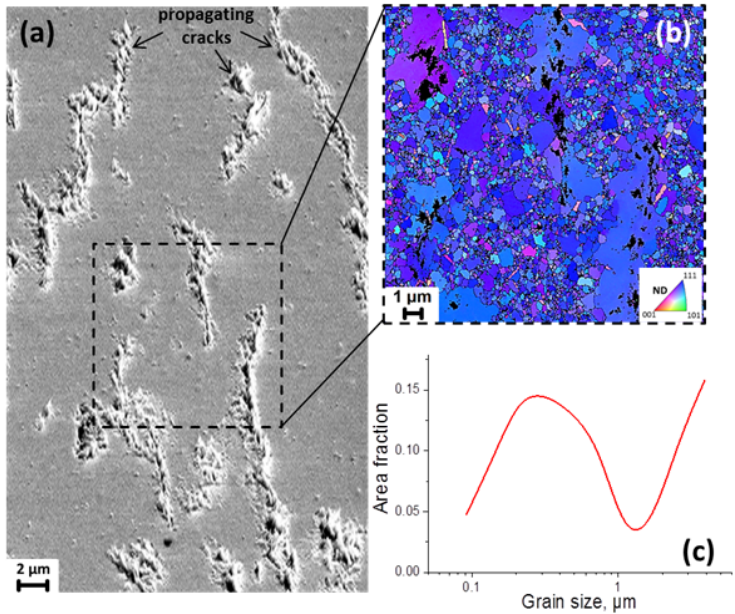

Fig. 8. The SEM image (a), grain orientation map (b), and corresponding grain size distribution (c) of 250 nm thick gold film after 70000 cycles with 1% cyclic strain. Increasing slop at the higher end of the grain size distribution (c) corresponds to the few heavily coarsened grains in (b). The straining direction is horizontal.

### 4.3. Discussion of the atomistic mechanism of grain coarsening.

The atomistic structure of grain boundaries cannot be resolved in the framework of the present EBSD-based experimental approach. Nevertheless, we can discuss how the available models for atomistic mechanisms of GB migration can be applied to explain the presented grain coarsening effect. As was shown in numerous TEM investigations [8,10,12,13,38], GB migration occurs by propagation of atomic-sized steps along the boundary. A simple sketch shown in Fig. 9 visualizes this mechanism. If two grains are separated by a GB with a step then lateral movement of this step will result in displacement of the whole GB. Direct experimental evidence of this mechanism was shown in the HRTEM study of room temperature migration of a grain boundary between [001] and [011] oriented grains in a gold film [38].

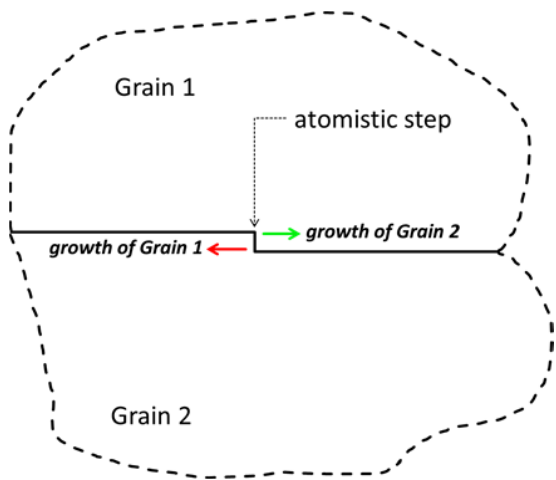

Fig. 9. General schematic of grain boundary migration through the movement of grain boundary steps.

For continuous migration of a grain boundary new steps must be generated, i.e. there must be a source of GB steps. It was demonstrated in [13] that a GB step can be generated by absorption of a lattice dislocation. In this way lattice dislocations can "feed" grain boundaries with new steps and support continuous GB migration. Such a mechanism is fully consistent with the thermodynamic driving force model presented here. One can assume that a part of the lattice dislocations generated by cyclic strain within the individual grains is consumed by the grain boundary. This results in, on the one hand, continuous generation of GB steps required for GB migration, and on the other hand, the absence of surface damage during the initial stage of grain coarsening as observed in Fig. 3.

Although the role of GB steps as mediators for GB migration seems to be well established, there is still no consensus regarding the necessity of coupling between the migration of a GB and shear displacement of the lattice behind the migrating boundary. Experimental demonstrations of shear coupled GB motion are restricted to rather specific cases. In the vast majority of reports shear coupling was observed at high homologous temperatures of more than $0.6T_m$ [7,11–13,15–17] for clearly defined specific GBs [7,12,13,15–17]. Thus, experimental evidences of shear coupled GB migration in polycrystals with general high-angle boundaries at room temperature are still missing.

Due to the insufficient resolution of the EBSD technique we can neither prove nor disprove the existence of shear coupling on the atomistic scale. However, we believe that shear coupling does not play a significant role in the observed grain coarsening due to the following reason. According to [12,13] the direction of shear-coupled GB migration depends on the Burgers vector of the disconnection and on the direction of the local shear stress which drives the movement of this disconnection. Thus, the shear coupled GB migration formalism cannot explain why virtually all GBs in Fig. 3a migrate exactly in the way which leads to isotropic growth of initially larger grains and final disappearance of UFG matrix (Fig. 3d). It is necessary to note that the thermodynamic driving force model proposed here is, in general, not in controversy with shear coupling formalism. The main difference is that our model is based on elastic strain energy density – a scalar quantity which is not directly related to the existence of a shear stress imposed on a GB. It can be therefore concluded that shear stress is not a necessary requirement for grain boundary migration if a sufficient thermodynamic driving force exists. To support this statement one can mention the texture transition effect leading to the growth of (100) grains at the expense of (111) grains in Cu and Ag thin films at relatively low homologous temperatures [39,40]. Although the exact nature of the driving force is still under discussion [40], it is a clear example of GB migration and selective grain growth due to a thermodynamic driving force. Another prominent example of grain growth induced by a specific driving force is texture transition under ion beam irradiation. It was shown that the grains with (110) orientation parallel to the ion beam grow significantly and can consume virtually all other grains in the film microstructure [41,42]. The driving force appears due to the large difference in defect density induced by ion bombardment since (110) orientation exhibit highest channeling of incident ions.

Thus, it is believed that GB migration and grain coarsening observed in the present work occurs predominantly through local shuffling of atoms across the boundary under the action of thermodynamic driving force. No shear coupling and even no resolved shear stress are necessary for the grain coarsening. This conclusion coincides with the last conclusion of the work of Babcock and Baluffi [10] who stated : "… the migration of high-angle grain boundaries, in general, occurs mainly by the local conservative shuffling of atoms across the boundary, either at pure steps or at other favorable places depending upon the boundary type."

## 5. Summary and conclusions.

A systematic study of room-temperature grain coarsening in thin gold films induced by cyclic strain is presented for three different film thicknesses and three different strain amplitudes. The quasi-in-situ EBSD data displays the microstructural evolution of a selected area of a film with increasing cycle number in detail. Simultaneously, the global grain size evolution as a function of strain cycles was captured by the significant statistical data analyzing more than 1000 grains for different numbers of cycles at different strain amplitudes. The presented results lead to the following conclusions:

1. Severe grain coarsening is induced in initially UFG gold films by application of cyclic strain at room temperature. The average grain size can be increased by an order of magnitude.

2. The grain coarsening is explained by a general thermodynamic driving force model which predicts the growth of the grains which have lower individual yield stress in comparison to surrounding matrix.

3. The main manifestation of the thermodynamic driving force model is reflected in a preferable isotropic growth of initially larger grains irrespectively of their crystallographic orientation or grain boundary misorientation angle.

4. Two grain coarsening stagnation mechanisms are described, both in correspondence with the thermodynamic driving force model. The first mechanism predicts that grain coarsening stagnates when a topological inversion from "soft grains in a hard matrix" to "hard islands in a soft matrix" occurs. The driving force disappears because the dislocation slip plasticity within the coarsened matrix becomes energetically more favorable. The second mechanism is based on the fact that propagation of through-thickness cracks perpendicularly to the applied strain results in local stress relaxation. Between the cracks, where the stress relaxation zones overlap, the driving force for grain coarsening is reduced due to the reduced local stress.

On the basis of the presented experimental results and analysis the following multi-step mechanism can be suggested to explain strain-induced room temperature grain coarsening. With increasing cycle number selected individual grains start to yield through dislocation glide. Microscopic yielding of individual grains results in a lower elastic strain energy density within these grains in comparison to the matrix. The difference in the stored elastic strain energy density constitutes the driving force for grain boundary migration. It is believed that a part of the lattice dislocations generated during each straining cycle is absorbed by the grain boundaries leading to permanent generation of grain boundary steps. The steps move according to the thermodynamic driving force, i.e. so that the yielded grains with lower strain energy density can grow in expense of the surrounding grains. This process continues with increasing cycle number until the driving force disappears according to the stagnation mechanisms described above.

Although the thermodynamic driving force model explains well the experimental observations presented in the current work, it is still not clear how widely it can be applied to different materials and different initial microstructures. Of particular interest is the question whether the strain-induced grain coarsening is restricted to highly textured films with ultra-fine grains or it can occur also in nanocrystalline and fine-grained materials in a similar way.

Acknowledgements. OG acknowledges full financial support from Austrian Science Fund (FWF) through the project P27432-N20. The authors would like to acknowledge the help of Rafael Soler in proving the reproducibility of EBSD data.